\documentstyle{article}
\begin{document}

\title{{\bf The Heumann-H\"otzel Model for aging revisited}} 
\author{Nazareno G. F. de Medeiros\thanks{Eletronic address:ngfm@
ifsc.sc.usp.br} and
Roberto N. Onody\thanks{Eletronic address:onody@ifsc.sc.usp.br}
\\
{\small {\em Departamento de F\'{\i}sica e Inform\'{a}tica } }\\
{\small {\em Instituto de F\'{\i}sica de S\~{a}o Carlos} }\\
{\small {\em Universidade de S\~{a}o Paulo - Caixa Postal 369} }\\ 
{\small {\em 13560-970 - S\~{a}o Carlos, S\~{a}o Paulo, Brasil.}}}
\date{} 
\maketitle 
\normalsize 
\baselineskip=16pt

\begin{abstract}

Since its proposition in 1995, the Heumann-H\"otzel model has remained as an 
obscure model of biological aging. The main arguments used against it were its 
apparent inability to describe populations with many age intervals and its 
failure to prevent a population extinction when only deleterious mutations are 
present. We find that with a simple and minor change in the model these 
difficulties can be surmounted. Our numerical simulations show a plethora
of interesting features: the catastrophic senescence, the  Gompertz law and that
postponing the reproduction increases the survival probability, as has already
been experimentally confirmed for the Drosophila fly. 

\vspace{1.5cm}


\vspace{0.5cm}


\end{abstract}

\newpage

\section{\bf Introduction}

\indent

Death is inevitable. It is usually preceded by a progressive deterioration
of our bodies. This phenomenon is called aging or senescence and it is
characterized by a decline in the physical capabilities of the
individuals. Although rare, some old people gazed at senescence with fine 
humour: "Old age is not so bad when you consider the alternative" said M. 
Chevalier (French singer and actor); "It is good to be here. At 98, it is good
to be anywhere" taught us G. Burns (US comedian and actor).

The new millennium, which is just beginning, will certainly be witness of a 
holy cruzade against aging. The principal
battle will be fought in the biochemical and medicine fields. Can physicist
help in any way? If we look at the progress made in the last decade,
we believe that the answer is yes. Indeed, physicists have brought new 
perpectives on the subject - the Occam's Razor principle. William of Occam,
a Franciscan monk, philosopher and political writer who was born in England in 
the thirteenth century, believed that for every phenomena occurring in the 
universe we need to look at the simplest explanation first - complexity should
not be assumed without necessity. This is the way physicists like to think of
nature but this is not followed by biologists. They love to see differences
and complexity where physicists love to see similarities and simplicity.
A good model in physics means one with a small number of parameters. With
the Occam's Razor principle in mind, what kind of aging model can we propose?

There are two kinds of aging theories: biochemical and evolutionary. 
The first invokes damages in cells, tissues and organs, the existence of free 
radicals or the telomeric shortening, that is, it sees senescence as a natural
consequence of biochemical processes \cite{kirk,red}. 
The second is the evolutionary theory \cite{rose,worth}, which 
explains the senescence as a competitive result of the reproductive rate,
mutation, heredity and natural selection.   

Evolutionary theories of aging are hypothetico-deductive in character,
not inductive. They do not contain any specific genetic parameter, but only
physiological factors and constraints imposed by the environment. There are 
two types: the optimality theory and the mutational theory. In the optimality 
theory \cite{topt}, senescence is a result of searching an optimal life 
history where
survival late in life is sacrificed for the sake of early reproduction.
A typical representative of such theories is the Partridge-Barton model
\cite{pb,ono1}.
For the mutational theory \cite{worth,tmut}, on the other hand, aging is a
process which comes from a balance between Darwinian selection and
accumulation of mutations. The natural selection efficiency to remove harmful
alleles in a population depends on when in the lifespan they come to express. 
Alleles responsible for lethal diseases that express late in life, escape
from the natural selection and accumulate in the population, provoking  
senescence. However, if the natural selection is too strong then
deleterious mutations might not accumulate \cite{ono2}. The most successful
aging theory of the mutational type is the Penna model \cite{pen,moss}.
By the way, throughout this paper, aging simply means that the average survival
probability of the population decreases with the age.

Here, in this paper, we analyse the Heumann-H\"otzel model \cite{hh}. Although
released at the same year as the Penna model it has remained in limbo. The
Achilles' heel of the Heumann-H\"otzel model was its incapacity to treat
populations with many age intervals (which all we expect to be a free
parameter in a reasonable model). Last but not least, in its original 
formulation the model could not handle mutations exclusively deleterious
(harmful mutations are hundred times more frequent than the beneficial ones)
leading to population meltdown. With minor modifications we were able
not only to repair those points but also to find some nice characteristics of
the model: it is Gompertzian, it exhibits catastrophic senescence and the effect
"later is better" (explained in the paper) is present.

\section{\bf The Heumann-H\"otzel Model}

\indent

In 1994, Dasgupta \cite{dg} proposed an aging model very similar to the 
Partridge-Barton \cite{pb} model, but without the antagonistic pleiotropy.
The antagonistic pleiotropy \cite{antg} arises when the same gene is responsible
for multiple effects. For example, genes enhancing early survival by promotion
of bone hardening might reduce later survival by promoting arterial hardening.
Reproduction is asexual.
As in the Partridge-Barton model, every individual in the Dasgupta model 
can have only three ages. 

Heumann and H\"otzel \cite{hh} generalized the Dasgupta model to support an 
arbitrary number of ages. However, when they simulated a population with eleven 
ages, they found that (in the final stationary state) there is again only 
three ages, recovering the Partridge-Barton results. This fact put the
Heumann-H\"otzel model in limbo. We will show later how some simple 
modifications can change drastically this scenario.

Let us now briefly describe the Heumann-H\"otzel model. At time $t$, there is
a population composed by $N(x,t)$ individuals with age $x$, $x=0 (babies),1,2,
...,$ $x_{max}$. Each individual carries a "chronological genome" of size $x_{max}$
with a survival probability per time step $G(x)$ at age $x$. 
There will be senescence if this
genome, averaged over the whole population, has $G(x)$ diminishing with $x$. At 
each time step $t$, {\it every} individual passes through the following stages:
\begin{itemize}
\item The Verhulst factor $V(t)$

The Verhulst factor plays the role of the environment (e.g., food restrictions).
It is given by
\begin{equation}
V(t) = 1 - \frac{N(t)}{N_{max}}
\end{equation}
where $N(t)$ is the total population at time $t$ and $N_{max}$ is a chosen
parameter. If an individual at age $x$ has $G(x) > 1 - V(t)$ then he survives 
to the
next step, otherwise he is eliminated. Actually, it is the Verhulst factor
which prevents the population to blow up.

\item The natural selection

A random number $r \in [0,1] $ is drawn. If an individual has $G(x) > r$ then
he survives.

\item The asexual reproduction

In the interval $R_{min} \le x \le R_{max}$ an individual has $m$ offsprings,
every one carrying a genome inherited from his father.

\item The mutations

At a randomly chosen position $x$, each individual has his survival probability
$G(x)$ mutated to $G'(x)$ by a random number $u$ ($-|a| \le u \le |b|$,
i. e., the mutations can be deleterious or beneficial)

\begin{equation}
G'(x) = G(x) e^{u}
\end{equation}

\item All individuals die after $x_{max}$.

\end{itemize}

As we said before, such a dynamics has two bad consequences: $x_{max} = 2$ and
there is population meltdown when only deleterious mutations are allowed. 

To overcome these difficulties, we made two simple but essential modifications:
mutations are allowed only on a fraction $F$ ($ 0 \le F \le 1$) of the {\it
babies} ($x=0$). By restricting the mutations only to the babies we brought the
model more close to reality, since it is well known that hereditary mutations
mostly take place during the reproduction process acting on the babies not
in their fathers. Mutations affecting the adults are predominantly of the
somatic kind. The original formulation - 
mutations happening for every
individual at any age - not only imposes a colossal rhythm of mutations
but also it seems to be unnatural. Furthermore, as there is a chance that
some babies could escape from mutations, we introduced the parameter $F$ which
represents the fraction of the mutated babies. In the presence of one 
parameter $F <  1$ there will be 
no population meltdown. We point out that such a parameter also
exist in the Penna model, but in an underground way.

Taking into account such modifications, we have performed some numerical 
simulations and found a lot of new and interesting results.

\section{\bf Our Results}

\indent

We simulated the modified Heumann-H\"otzel model with the starting condition:
$N(x,t=0)=N_{0} \delta_{x,0}$. The initial number of babies, $N_{0}$, varied 
from 10 to 20,000 and the genome distribution was chosen to be random or "pure",
which means that all babies have the genome $G(x)=1, \; \forall x$. For $N_{0}=
20,000$, we didn't find any qualitative difference between the two 
distributions. Of course, for very small initial population only the pure 
distribution can reach the stationary regime. 
In general, we run 300,000 time steps and average all quantities over the last
10,000 steps (when the stationary regime had already been achieved). 
We fixed $N_{max}=800,000$. The measured quantities were:
$N(x,t)$, $\langle N(x) \rangle$ - the time average over $N(x,t)$,
$\langle G(x,t) \rangle$ - the average genome of the individuals with
age $x$ at the instant $t$, $\langle G(x) \rangle$ - the survival 
probability - i. e., the time average over $\langle G(x,t) \rangle$. 

Figure 1 (a) shows that the modified Heumann-H\"otzel model does not lead to a
population meltdown even when the mutations are exclusively 
deleterious. Figure 1 (b) indicates that the stationary regime has been 
achieved after $t=180,000$.

Some recent experiments \cite{pf}, done with the fruit fly drosophila 
melanogaster, unequivocally demonstrate that postponing reproduction favours 
the population. In the experiments, the male and female flies were put 
together some time later than they have reached their sexual maturity. 
The result
was an improvement of the population characteristics - both male and female
flies have increased their survival probabilities at old ages. Ok, the
Heumann-H\"otzel model treated here is asexual, but we can think of this effect
- later is better - by studying what happens if we delay the initial 
reproduction age $R_{min}$. Figure 2 shows that the effect "later is better" is
in fact present.

While iteroparous individuals can breed repeatedly, semelparous individuals
breed only once. The Pacific salmon is a good example of the later. This fish
has a dramatic manifestation of aging, the so called catastrophic senescence.
It dies just after its sexual maturity. The Heumann-H\"otzel model exhibits
the catastrophic senescence if we make $R_{min}=R_{max}$ (see figure 3).

As a final study, we verified that the Heumann-H\"otzel model obeys nicely the
Gompertz law.
Based on actuarial observations, Gompertz \cite{go} found in 1825 that the
human mortality function 

\begin{equation}
q(x) = - \frac{d \ln \langle N(x) \rangle}{dx}
\end{equation}
grows up exponentially with the age $x$ for some interval age. 
Removing the Verhulst factor, i. e.,
considering only deaths by natural causes, it is easy to show that
$q(x) = 1 - \langle G(x) \rangle $.

Figure 4 shows our result. The Gompertz 
law is satisfied only in an age interval (see discussion below).

\section{\bf Discussion}

\indent

In this paper, we have modified the Heumann-H\"otzel model in order to fix
its main deficiency - to be a senescence model with only three possible ages.   
The crucial point was to change the huge amount of mutations. This was done by
permitting only mutations in a fraction $F$ of the babies. The consequences
were surprisingly very good. 

The first achievement was to get an arbitrary (and stable) number of ages.
Without this would be impossible to obtain all the other results. Our numerical
data show that the modified Heumann-H\"otzel model has some sensitive and also
robust parameters. From a qualitative point of view, it does'nt matter what
type of initial conditon we use: random or "pure". The two parameters:
$N_{max}$ (of the Verhulst factor) and $m$ (fertility) only determine the 
size of the final stationary population. We tried to implement the Verhulst
factor ($V(t)$) in two different ways: i) for each individual (with any age 
$x$) a random number $r$ is drawn and if $ r < 1 - V(t) $ he dies (aleatory
decimation); ii) each individual at age $x$ with $ G(x) < 1 - V(t) $ dies 
(discriminatory decimation). We did not detect any important differences 
between these two cases. On the other hand, the three parameters: $F$
(fraction of mutated babies), $a$ (deleterious mutation intensity) and $b$
(beneficial mutation intensity) are very sensitive: a bad choice may lead
to a population meltdown. This is what happens for example for $b=0$, $F=1$
and arbitrary $a$. Finally, the other two parameters: $R_{min}$ and $R_{max}$
(corresponding to the reproductive age interval) may also conduct to a 
population meltdown if not wisely chosen. Postponing the sexual maturity, 
i. e., increasing $R_{min}$ (up to a maximum value over which there will be
again extinction of the population) clearly favours aging, in the sense that
now the individuals live longer. Although, as far as we know, such a 
phenomenon has only been detected in organisms with sexual reproduction,
it was amazing to find it also here. If $R_{min}=R_{max}$, i. e., if an
individual reproduces only once in his entire life, there will be a 
catastrophic senescence.

In figure 5 we show the mortality function in the year 1998 for Brazil and
USA. The Gompertz law holds only in an age interval, something between the
ages 35-60. It is only in this region that the mortality function grows up
exponentially. Our simulation (figure 4) exhibits the same behavior in the
interval 4-9. Here again, sex does not seem to play any relevant role.
Comparing the two plots one can see that the modified Heumann-H\"otzel
model does not reproduce correctly neither the high infantile mortality 
(figure 5, for $x < 10$) nor the mortality decayment for $x > 80$ (not
shown in the figure 5). For the fruit fly Drosophila such a late-life
decayment forms a plateau \cite{mue}. On the other hand, the modified 
Heumann-H\"otzel model predicts (see figure 4) that the mortality is 
almost constant for $x < 4$ and it grows up faster than an exponential
for $x > 10$. These discrepancies are almost certainly connected to the
fact that, independently of their ages, the Verhulst factor (which plays the 
role of the environment) is the same for all individuals. So a Verhulst
factor varying with the age $x$ would be very nice.  
Recently \cite{pol}, it was found a kind of infantile mortality by defining
the Penna model in a lattice.

To conclude, let us compare the Heumann-H\"otzel model with the Penna model.
There are some pros and cons. One difference is how the natural selection 
is implemented in each model. In the Penna model there is a parameter
(fixed and equal for all individuals) - the threshold - which is a kind of
destiny. Any individual having a number of mutations greater than the
threshold dies. The Heumann-H\"otzel model on the other hand gives to any
individual a chance to escape from natural selection even when his survival
probability is very small. This aleatory aspect is more close to the 
Darwinian ideas of natural selection. In the Penna model a
deleterious mutation cannot occur twice in the same position of the 
genome. This means that some babies will not mutate. Through the parameter $F$
this feature is also present in the modified Heumann-H\"otzel model, but here
the number of deleterious mutations in the same position of the genome is
not restricted or limited. Finally, due to its bit-string characteristics, 
the Penna model is easier to simulate and it is more suited to 
treat large populations. The Heumann-H\"otzel model, on the other hand, can 
start with a very small population ($N_{0} \sim 10$ of "pure" babies) and still 
reaches the stationary regime. In a conversation with Prof. Stauffer he
recommended to include sex in the Heumann-H\"otzel model. This is now under way.

\section {\bf Acknowledgements}
 
\indent

We acknowledge CNPq (Conselho Nacional de Desenvolvimento Cient\'{\i}fico e 
Tecnol\'ogico) and FAPESP (Funda\c c\~ao de Amparo a Pesquisa do Estado de
S\~ao Paulo) for the financial support.

\newpage

\begin{center} 
FIGURE CAPTION 
\end{center}

\vspace{1.5cm}

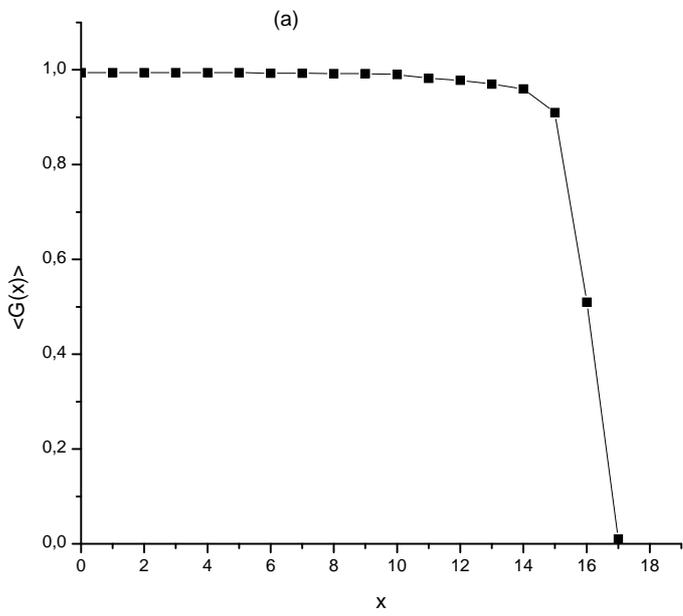
Figure 1. The Heumann-H\"otzel model with 18 ages. The parameters used were:
$F=0.1$, $m=1$, $|a|=0.2$, $|b|=0$, $R_{min}=8$ and $R_{max}=17$. (a) plot 
of the survival probability versus the age; (b) time evolution of the number
of individuals with age $x$, $N(x,t)$. From top to bottom $x=0,...,17$. 

\vspace{1.5cm}

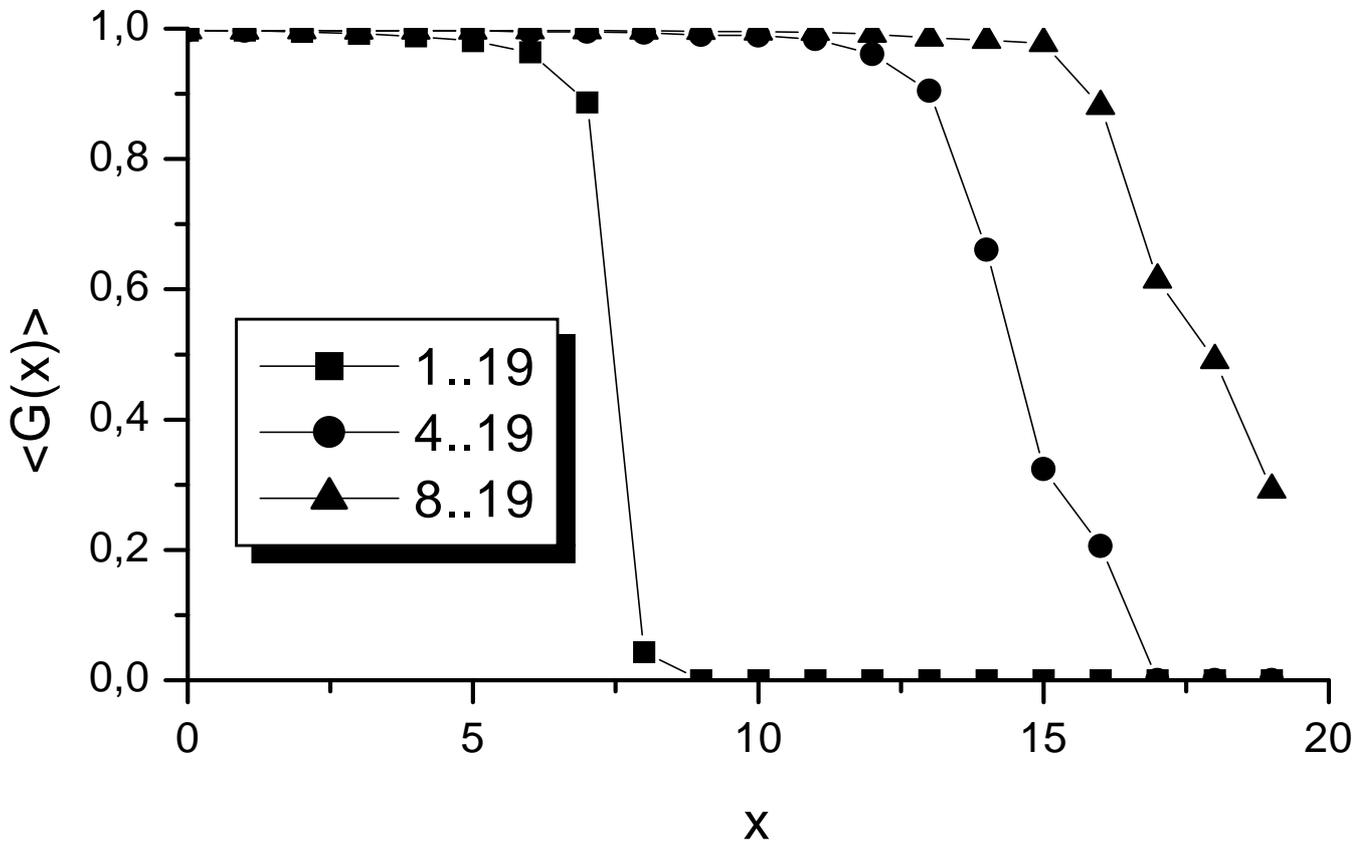
Figure 2. The effect "later is better" in the modified Heumann-H\"otzel model.
The parameter's values are: $F=0.1$, $m=1$, $|a|=0.04$ and $|b|=0.02$. The 
interval age of reproduction $R_{min}..R_{max}$ corresponding to the curves
are shown in the inset.

\vspace{1.5cm}

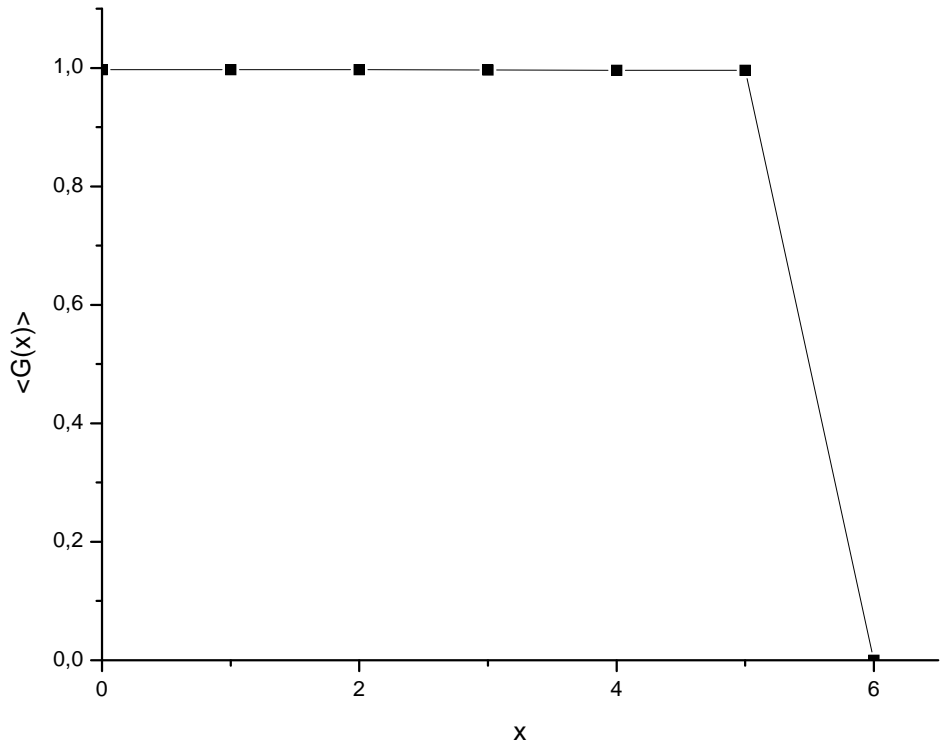
Figure 3. The catastrophic senescence in the modified Heumann-H\"otzel model.
The parameter's values are: $F=0.1$, $m=4$, $|a|=0.04$, $|b|=0.02$ and
$R_{min}=R_{max}=5$.

\vspace{1.5cm}

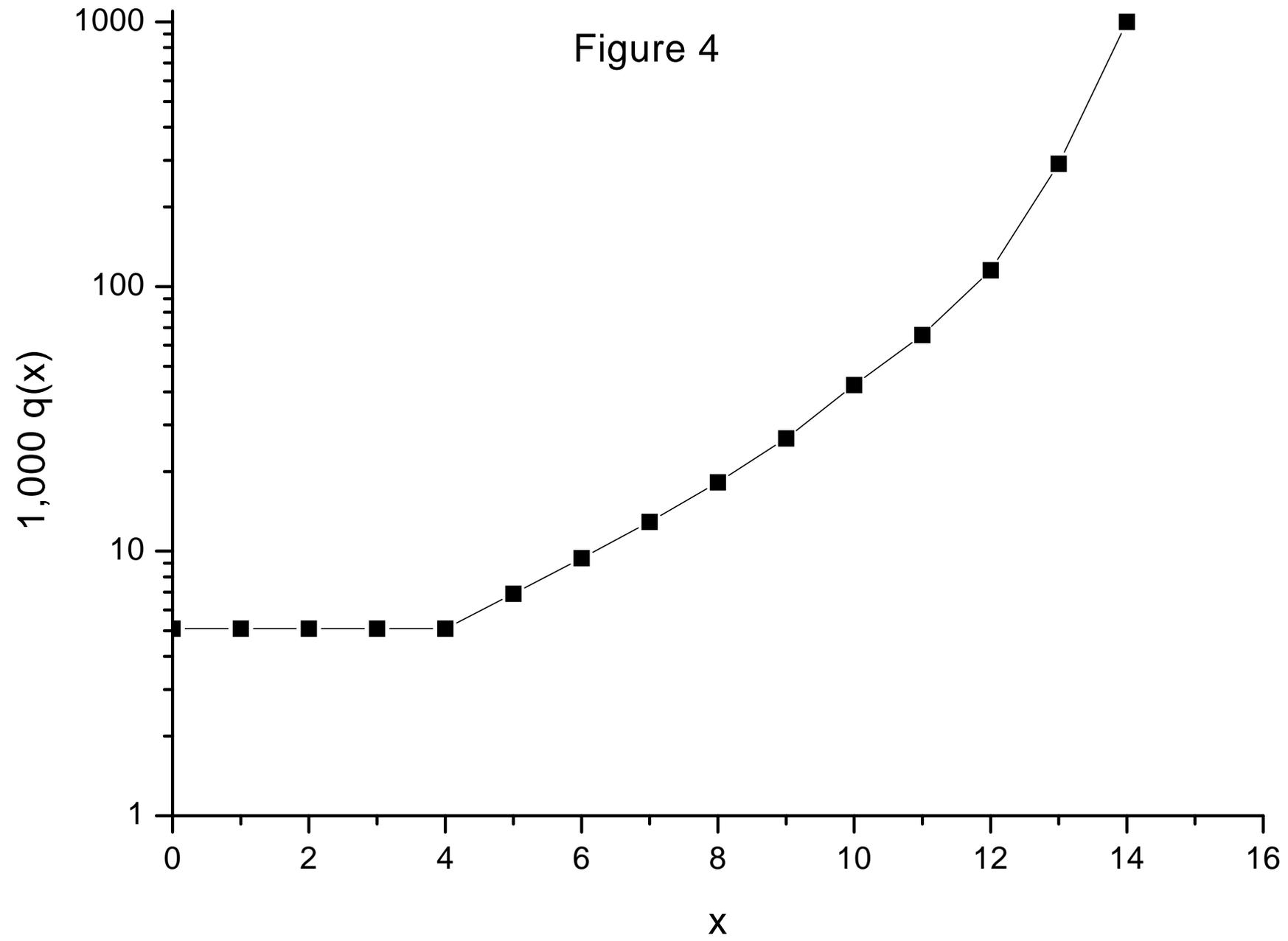
Figure 4. The Gompertz law holds in the age interval
$ x \in [4,9]$. The parameters are: $F=0.1$, $m=1$, $|a|=2$, $|b|=0$, 
$R_{min}=4$, $R_{max}=15$.

\vspace{1.5cm}

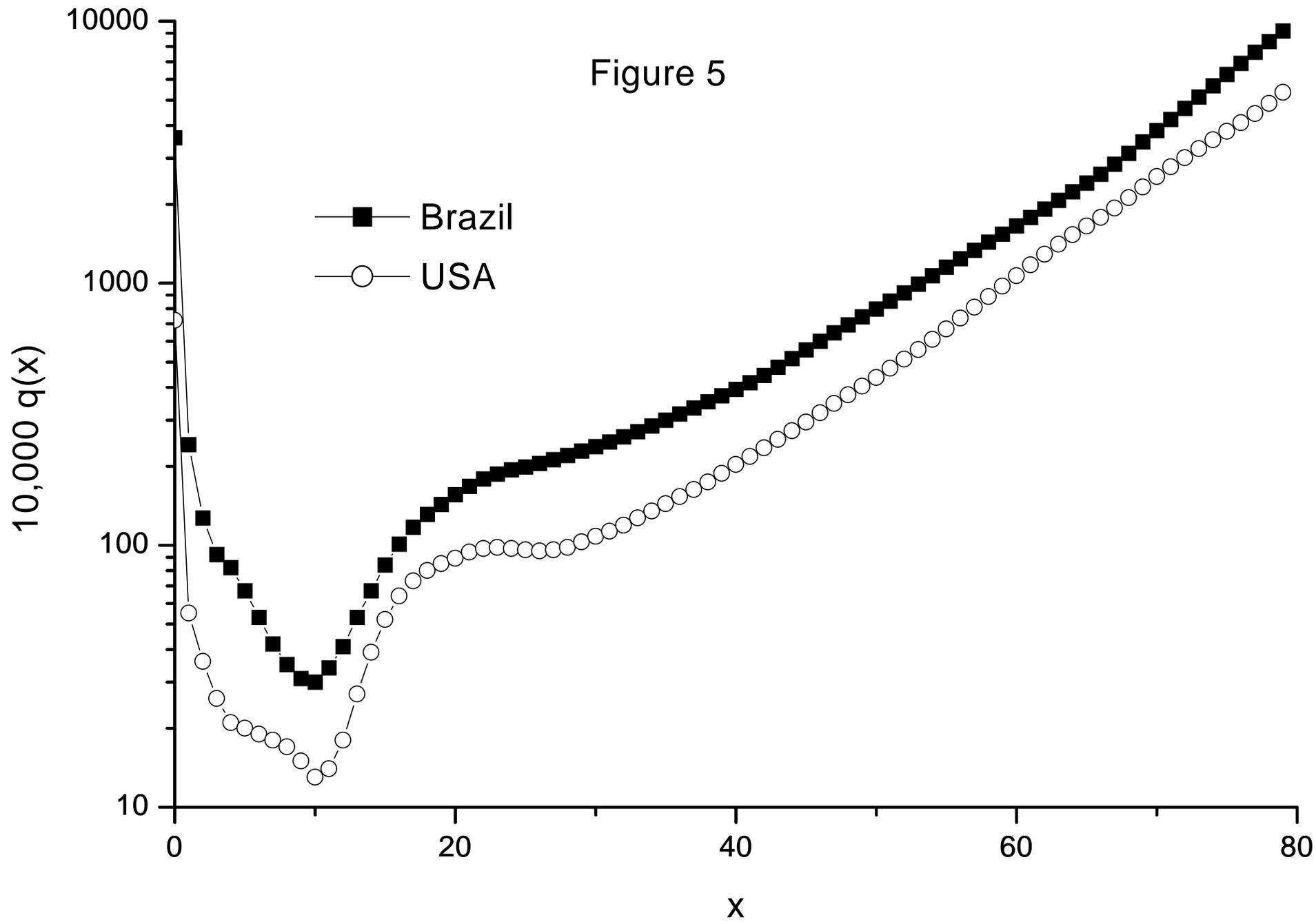
Figure 5. Mortality in Brazil and USA in the year 1998. Searches: Instituto 
Brasileiro de Geografia e Estat\'{\i}stica and U. S. Department of Health and
Human Services. The Gompertz law holds in the age interval $ x \in [35,60]$.


\newpage

\begin{thebibliography}{99}

\bibitem{kirk} Kowald A. and Kirkwood T. B. L., J. Theor. Biol. {\bf 168},
75 (1994).

\bibitem{red}  Reddel R. R., BioEssays {\bf 20}, 977 (1998).

\bibitem{rose} Rose M. R., {\it Evolutionary Theory of Aging}, Oxford 
University Press, Oxford, 1991.

\bibitem{worth} Charlesworth B., {\it Evolution in Age-Structured Populations},
Cambridge Univ. Press, 1994.

\bibitem{topt} Parker G. A. and Maynard Smith J., Nature {\bf 348}, 27 (1990).

\bibitem{pb} Partridge L. and Barton N. H, Nature {\bf 362}, 305 (1993).

\bibitem{ono1} Onody R. N. and de Medeiros N. G. F., Phys. Rev. E, {\bf 61},
5664 (2000).

\bibitem{tmut} Medawar P. B., {\it An Unsolved Problem in Biology}, Lewis, 
London, (1952).

\bibitem{ono2} Onody R. N. and de Medeiros N. G. F., Phys. Rev. E, {\bf 60},
3234 (1999).

\bibitem{pen} Penna T. J. P., J. Stat. Phys., {\bf 78}, 1629 (1995).

\bibitem{moss} Moss de Oliveira S., de Oliveira P. M. C. and Stauffer D.,
{\it Evolution, Money, War and Computers}, Teubner, Leipzig (1999).

\bibitem{hh} Heumann M. and H\"otzel M., J. Stat. Phys. {\bf 79}, 483 (1995).

\bibitem{dg} Dasgupta S., J. Phys. I (Paris) {\bf 10}, 1563 (1994).

\bibitem{antg} Rose M. R., Heredity {\bf 48}, 63 (1982); Rose M. R., 
Theor. Pop. Biol. {\bf 28}, 342 (1985); Rose M. R., {\it Genetic Constraints on
Adaptive Evolution} (ed. Loeschcke, V., Springer, Berlin), 91 (1987).

\bibitem{pf} Partridge L. and Fowler K., Evolution {\bf 46}, 76 (1992).

\bibitem{go} Gompertz B., Phil. Trans. Roy. Soc. London {\bf A 115}, 513 (1825).

\bibitem{mue} Mueller L. D. and Rose M. R., Proc. Natl. Acad. Sci. USA {\bf 93}, 
15249 (1996).

\bibitem{pol} Makowiec D., Physica A {\bf 289}, 208 (2001).

\end{thebibliography}
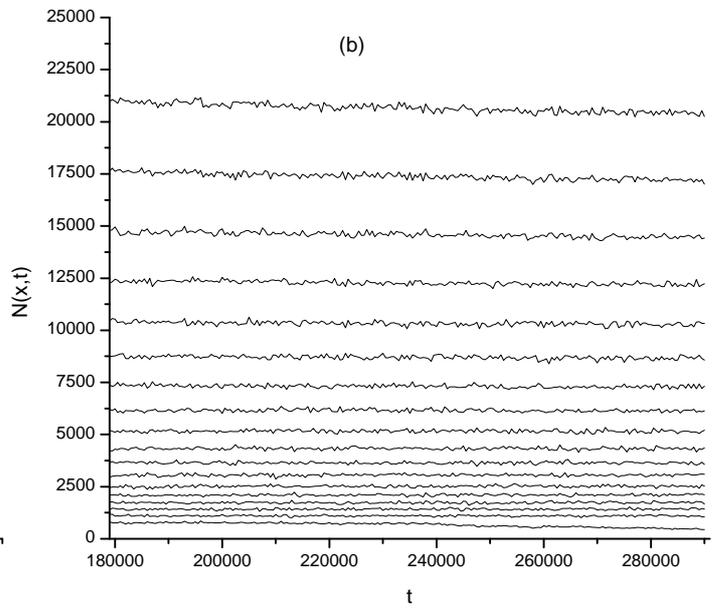
\end{document}